# Enhancing Swelling Kinetics of pNIPAM Lyogels: The Role of Crosslinking, Copolymerization, and Solvent

Kathrin Marina Eckert[a]*, Jelisa Bonsen[a], Anja Hajnal[a], Johannes Gmeiner[b], Jonah Hasse[b], Muhammad Adrian[c], Julian Karsten[d], Patrick A. Kißling[e], Alexander Penn[c], Bodo Fiedler[d], Gerrit A. Luinstra[b], Irina Smirnova[a]

* corresponding author (kathrin.eckert@tuhh.de)

*[a] Institute of Thermal Separation Processes, Hamburg University of Technology, Hamburg, Germany*

*[b] Institute for Technical and Macromolecular Chemistry, University of Hamburg, Hamburg, Germany*

*[c] Institute of Process Imaging, Hamburg University of Technology, Hamburg, Germany*

*[d] Institute of Polymers and Composites, Hamburg University of Technology, Hamburg, Germany*

*[e] Institute of Chemical Reaction Engineering, Hamburg University of Technology, Hamburg, Germany*

## ABSTRACT

Stimuli-responsive lyogels are known for their ability to undergo significant macroscopic changes when exposed to external stimuli. While thermo-responsive gels, such as poly-N-isopropylacrylamide (pNIPAM), have been extensively studied across various applications, solvent-induced swelling has predominantly been investigated in aqueous solutions. This study explores the tailoring of lyogel formulations for future applications by controlling their solvent-induced swelling behavior, comparing both homopolymeric and semi-interpenetrating polymer networks (semi-IPNs). It is structured in two parts: the first focuses on characterization techniques, including NMR relaxometry, swelling degree measurements, mechanical testing, and SEM analysis, while the second part delves into swelling kinetic analysis, applying solvent exchange as a stimulus for varying gel formulations and solvents. In contrast to most previous studies, the impact of chemical and physical crosslinking, as well as copolymer inclusion, on the swelling behavior and mechanical properties of lyogels in organic solvents is examined and compared with solvent-induced swelling kinetics measurements. The results demonstrate that increasing chemically crosslinking in homopolymers and physically crosslinking in semi-IPNs enhances mechanical stability, while improving mass transport properties and solvent exchange kinetics. However, increases degree of crosslinking results in a prolonged response time to the solvent exchange stimulus and a reduction in the overall swelling capacity of the lyogels. Furthermore, variations in solvent properties, including molecular size and diffusion rates, significantly influence the swelling kinetics, whereas smaller, faster-diffusing solvents leading to more pronounced solvent spillage effects. Our findings highlight the complex interplay between gel formulation, network structure, and solvent nature in determining the solvent-induced swelling kinetics of lyogels, providing insights into how these materials can be tailored for specific applications especially those requiring short response times and optimized mechanical properties.

## INTRODUCTION

Stimuli-responsive gels have the inherent capability to undergo large macroscopic changes when exposed to external stimuli[1], [2], [3], [4]. These stimuli induce solvent uptake or expulsion, resulting in volumetric changes in the gel structure[1], [2]. While temperature-[5], [6], [7] and pH-induced[8], [9], [10] swelling of hydrogels in water have been widely investigated, solvent-induced swelling[4], [11], [12], [13] has not been a primary focus of research. This special characteristic of stimuli-responsive gels enables their application in various fields, such as drug release[5], [8], [14] or reactor control[15], [16]. Among the most extensively studied stimuli-responsive polymers are poly-N-isopropylacrylamide (pNIPAM)[4], [17], [18], [19], [20], [21], polyacrylic acid (PAA)[22], [23], [24], or poly-N,N-dimethylaminoethyl methacrylate (DMAEMA)[8], [22]. However, most applications require specifically tailored materials to optimize their performance in the intended processes. In this work, we emphasize the impact of polymer formulation and solvent nature on the gel's swelling response to tailor polymers for potential applications beyond the established principles of hydrogel swelling in various solvent-induced swelling processes.

Solvent exchange and solvent-induced swelling processes of lyogels are not only important in the context of stimuli-responsive gels but also in other processes, as aerogel production[25], [26] or various adsorption processes[27], [28]. The swelling of hydrogels is well-studied and described for a broad range of polymers under various stimuli, including temperature[21], [29], [30], [31] or pH[9], [10], [32], [33], [34], [35]. Recent studies have also explored the swelling in different solvents[4], [12], [13], [36]. However, the swelling kinetics of gels are primarily described for hydrogel swelling[18], [23], [37], [38], [39], [40], [41], [42].

Since swelling is a defining characteristic of these gels, response kinetics often represent a critical limitation for many applications. To improve response kinetics and expand potential applications, the effects of copolymers and gel size have been widely explored in the literature[19], [30], [35], [41] and are further examined in this work. It is known that pNIPAM homopolymers respond relatively slowly to external stimuli, though the exact time scale may vary depending on factors such as type of stimulus and gel size. According to Haq *et al.*[19], this slow response is not due to the polymer type itself but rather to the three-dimensional structure of pNIPAM gels[19]. In contrast, semi-interpenetrating polymer networks (semi-IPNs) incorporating pNIPAM enable faster water transport into the matrix, resulting in improved kinetic responses[19], [30], [43]. Compared to hydrogel swelling, the kinetics and behavior of stimuli-responsive gels in organic solvents have mostly been analyzed in aqueous solutions[29], [44], [45], but rarely in pure organic solvents[4], [12], [13], [46]. In particular, solvent exchange kinetics in organic solvents have not been reported extensively in the literature, and are therefore the focus in this work.

Investigating the mechanisms of swelling and shrinkage, hydrogels typically follow an exponential or power-law trend, as described by Ritger[47], or Peters[38], shown in various experimental studies[18], [25], [41], [43], [48], [49]. For aerogel production, the solvent exchange kinetics from water to ethanol have been studied in detail, notably by Dirauf *et al.*[25]. Their findings revealed that the solvent exchange kinetics in the systems analyzed exhibit a trend similar to the swelling kinetics of hydrogels[25]. However, not all solvent exchange steps exhibit the same trend as that observed in hydrogel swelling[50], [51]. Bueno *et al.*[50] were the first to identify a solvent spillage effect during the solvent exchange from ethanol to supercritical carbon dioxide ($CO_2$) in aerogel production. This effect consists of three main stages: (1) $CO_2$ diffusion into the gel matrix: The internal pressure increases continuously. Capillary forces retain the liquid in the pores, but ethanol diffuses out of the matrix. (2) Advective spillage: Once the pressure threshold for retaining liquid in the pores is exceeded, an advective flow of ethanol is expulsed, along the continued ethanol diffusion out of and $CO_2$ diffusion into the matrix. (3) Continuing solvent spillage: The solvent spillage continues at slower rates, along with the diffusive flows of ethanol and $CO_2$. Compared to the solvent exchange kinetics of hydrogels, the advective solvent spillage introduces an overlaying factor, resulting in altered swelling kinetics and shrinkage in these systems[50]. Similar swelling kinetics have also been reported by Takeshita *et al.*[51].

In this study, the solvent-induced swelling mechanisms of pNIPAM are investigated in context with solvent spillage as well as established swelling and shrinkage mechanisms. The focus was on the influence of different solvents and copolymerization in tailoring lyogels for possible applications. We included solvents representing different functional groups (alcohols, carboxylic esters, and carboxylic acids) and molecule sizes, as well as an esterification system as a potential application. Additionally, the swelling kinetics of pNIPAM homopolymers are compared with those of pNIPAM-polyvinyl alcohol (PVA) semi-IPNs.

## EXPERIMENTAL METHODS

### Materials

The following chemicals and organic solvents were used as purchased: N-isopropylacrylamide (NIPAM) (> 98%), N,N'-methylenebis(acrylamide) (MBA) (> 99%), ammonium persulfate (> 98%), and sodium persulfate (> 97%), acetic acid (≥ 99.8%), 1-butanol (≥ 99.5%), butyl acetate (≥ 99.5%), ethanol (≥ 99.9%), ethyl acetate (≥ 99.7%), methanol (≥ 99.8%), methyl acetate (≥ 99.0%), polyvinyl alcohol (PVA) (≥ 99.9%) (Table S1).

**Gel Synthesis and Preparation**

For the preparation of hydrogels, 2.175 g NIPAM and the required MBA amount (according to Table 1) were dissolved in 21 g of deionized water, and degassed with nitrogen for 30 min. Subsequently, 2.5 mg of sodium persulfate and ammonium persulfate were each dissolved in 1 mL of deionized water and added to the mixture. The hydrogel monoliths were cast in syringes (inner diameter: $d_3$ = 1 mm, $d_5$ = 1.3 mm, $d_{10}$ =1.65 mm; BRAUN syringes of 3, 5, and 10 mL) and covered with PARAFILM (Sigma-Aldrich). For the PVA-containing gels, the 21 g of deionized water was replaced with a PVA-water mixture according to Table 1. In all formulations, 0.01 g of coloring pigment (Kremer Pigmente, Fluorescent Pigment Magenta Red) was added for the analysis of swelling equilibria and kinetics.

For analyzing the impact of freeze-thaw cycles on swelling behavior, the freeze-thaw cycle was performed by freezing the gels at -18 °C for 24 h, followed by thawing at room temperature for 24 h.

**Table 1. Composition of the gel formulations including crosslinker (MBA) and copolymer (PVA) concentrations.**

| | Formulation | Crosslinker concentration (MBA : NIPAM) [$g_{MBA}/g_{NIPAM}$] | Copolymer concentration (NIPAM : PVA) [$g_{NIPAM}/g_{PVA}$] |
|---|---|---|---|
| **Homopolymer** | pNIPAM-1 | 0.007 | - |
| | pNIPAM-2 / PVA-0 | 0.014 | - |
| | pNIPAM-3 | 0.020 | - |
| | pNIPAM-4 | 0.035 | - |
| | pNIPAM-5 | 0.050 | - |
| **Semi-IPN** | PVA-1 | 0.014 | 1.036 |
| | PVA-2 | 0.014 | 2.071 |
| | PVA-3 | 0.014 | 4.143 |

**Swelling Equilibria of Lyogels**

After synthesis, and potential freeze-thaw cycles, the hydrogel cylinders were cut into smaller cylindrical segments (height: h = 1-1.5 cm); discarding inhomogeneous end pieces (Figure S1). For the solvent exchange from water to organic solvents, the gel monoliths were immersed in the desired solvent and shaken, followed by two additional solvent exchange steps with fresh solvents each after 24 h. To study the effect of 1-butanol an additional solvent exchange to ethanol is required due to its limited miscibility with water. Subsequently, the lyogel was equilibrated with the corresponding solvent for 48 h at 25 °C (± 0.1 °C) in a water bath (Grant OLS200). The required equilibration time was experimentally determined beforehand. To calculate the mass-specific degree of swelling *Dos* (Eq. 1), the equilibrated lyogels were dried at 40 °C in a vacuum oven (VT 6060 M-BL, Thermo Scientific).

$$Dos = \frac{m_{equilibrated\ gel}}{m_{dried\ polymer}} \tag{1}$$

**NMR Relaxometry**

A Nuclear Magnetic Resonance (NMR) relaxometry analysis provides valuable insights into the evolution of the gel’s structure and properties during the solvent exchange process. Custom glass holders were designed and fabricated using Selective Laser-induced Etching (SLE), ensuring reproducible sample positioning (Figure S2). All NMR measurements were performed using a 60 MHz benchtop NMR spectrometer (Spinsolve 60, Magritek, Germany) with a 5 mm diameter NMR tube. To characterize the hydrogel samples, a CPMG (Carr-Purcell-Meiboom-Gill) pulse sequence was employed, using pulse length of 13.8 µs, an echo time of 300 µs, 50,000 number of echoes, and a repetition time of 20,000 ms, providing insights into the structure and dynamics of the gel. The CPMG sequence was utilized to measure transverse relaxation times $T_2$, providing information about molecular mobility and interactions within the gel matrix[52], [53], [54]. By analyzing the variation of $T_2$ relaxation times, it was possible to gain insights into the different molecular environments present within the gel structure. In this study, the analysis was performed using mono-exponential fitting of the $T_2$ decay. In the future, more sophisticated approaches, including multiexponential fitting of the $T_2$ decay (e.g. using the inverse Laplace transform), as described by the work von Demco and Pich[55], could be employed to obtain a more in-depth analysis of gel matrices.

**Mechanical Properties**

The mechanical properties of the hydrogels were evaluated through uniaxial compression tests using a universal testing machine (ZwickRoell GmbH & Co. KG, Universalprüfmaschine Z1474, nominal force: 2.5 kN) at ambient conditions. Prior to testing, the hydrogels were immersed in deionized (DI) water for at least 48 h. During the test, the compression stamp ($d_{stamp}$: 40 mm, unconfined compression test: $d_{stamp} > d_{sample}$) was initially advanced toward the gel until an initial force of 0.02 N was detected. Subsequently, the gel was compressed at a rate of 50 mm/min until either a maximum deformation of 75% or the maximum compression force of 50 N of the load cell (ZwickRoell GmbH & Co. KG, Xforce HP, nominal force: 50 N), was reached. The progression of the force over the crosshead travel was recorded throughout the test. The elastic modulus (Young's modulus $E_C$) was determined within the elastic deformation range of the compression curve at 10-12% deformation.

**Scanning Electron Microscopy Analysis**

For the scanning electron microscopy (SEM) analysis, a Zeiss Supra VP 55 equipped with a cold field emission gun and an in-lens detector was used for imaging, whereby each measurement was carried out under high vacuum using an acceleration voltage of 5.00 kV. Prior to analysis, in ethanol equilibrated samples were dried using supercritical $CO_2$. In this step, the alcogels were sealed in filter paper bags and transferred into a custom-built, high-pressure autoclave with a volume of 250 mL. The drying was performed at 60 °C and 125 bar for 3 hours with a continuous $CO_2$ flow. The dried samples were cut open and sputtered with a thin carbon layer (approx. 10 nm) (Sputter Coater SCD 050, Oerlikon Balzers AG, Balzers, Liechtenstein) prior to analysis.

It must be emphasized that the SEM analysis only reflects the structure of the supercritical dried gels. Therefore, no definitive conclusions about the pore structure in the wet gels can be drawn; only the expected trends in kinetic responses can be inferred from these data.

### Swelling Kinetics Analysis

The preparation of the monoliths for swelling kinetic analysis was conducted analogously to the procedure for swelling equilibrium measurements. The sliced hydrogels were solvent-exchanged with the selected solvents under study and subsequently equilibrated. In this study, the hydrogels were solvent-exchanged with ethanol, methanol, acetic acid, or 1-butanol. As previously described for the swelling equilibrium analysis, the solvent exchange to 1-butanol required a prior exchange step with ethanol. The solvents are selected because they occur in esterification reactions, and the analyzed solvent exchange kinetics can simulate the gel response during the progression of this reaction.

The lyogels were secured in place with a needle along their height and suspended in a custom-made wire mesh. Solvent-induced swelling was studied during the exchange from the pure solvent (alcohol or carboxylic acid) to its corresponding ester derivative. To achieve this, the gels were placed in a 250 mL DURAN beaker, filled with 200 mL of the ester. Changes in gel dimensions in response to ester exposure were recorded using a camera (Kurokesu C1 pro with 5-50 mm CS lens), capturing one image per minute for 24-48 h, depending on the solvent system.

The acquired image sequences were analyzed using a Python script, which detected the gel height and diameter by summing the number of pixels. From this analysis, volumetric changes were calculated.

For the dimensionless swelling kinetics, the time at which 50 % of the final volumetric shrinkage was reached was determined ($t_{50}$), while the dimensionless volume was referenced to the volume at the beginning of the experiment ($t$ = 0), denoted as $V_0$.

## RESULTS AND DISCUSSION

This study demonstrates the ability to tailor lyogel formulations to meet the requirements of potential applications. This is achieved by controlling the solvent-induced swelling behavior of lyogels, in contrast to prior studies that primarily focused on swelling in aqueous solutions[8], [30], [43], [56], [57], [58]. To enable a comprehensive analysis of their response characteristics, the lyogels are evaluated in terms of swelling capacity and mechanical properties. Based on these findings, the impact of gel formulation and solvent nature on the solvent-induced swelling behavior is analyzed.

**Characterization of Lyogels**

The gel formulations investigated in this study comprise homopolymers and semi-IPNs with varying degrees of crosslinking. In the homopolymeric formulations (pNIPAM 1–5), the crosslinker concentration is directly varied, while in the semi-IPN formulations (PVA 1–3), physical crosslinking is induced through freeze-thaw cycles at different PVA concentrations. The increased physical crosslinking during these cycles leads to the formation of semi-crystalline regions within the network [58], [59].

The increasing degree of crosslinking in the gel formulations influences both swelling behavior and mechanical stability (Figure 1). In all polymer formulations — homopolymers and semi-IPNs — higher crosslinking imposes greater restrictions of the polymer chains, thereby reducing the gels' swelling capacity (Figure 1a, b). The NMR relaxometry data reveal a corresponding decrease in $T_2$ relaxation times with increasing crosslinker concentration (Figure 1c, d) and decreasing swelling capacity[52], [53], [54] which may be attributed to stronger polymer-solvent interactions or smaller pore sizes. The lower solvent content in highly crosslinked gels, resulting from reduced swelling capacity, increases their mechanical stability, as indicated by increased Young's moduli (Figure 1e, f). This effect is observed for both chemically crosslinked homopolymers and physically crosslinked semi-IPNs, where freeze-thaw cycles provide additional network reinforcement (Figure 1b, f).

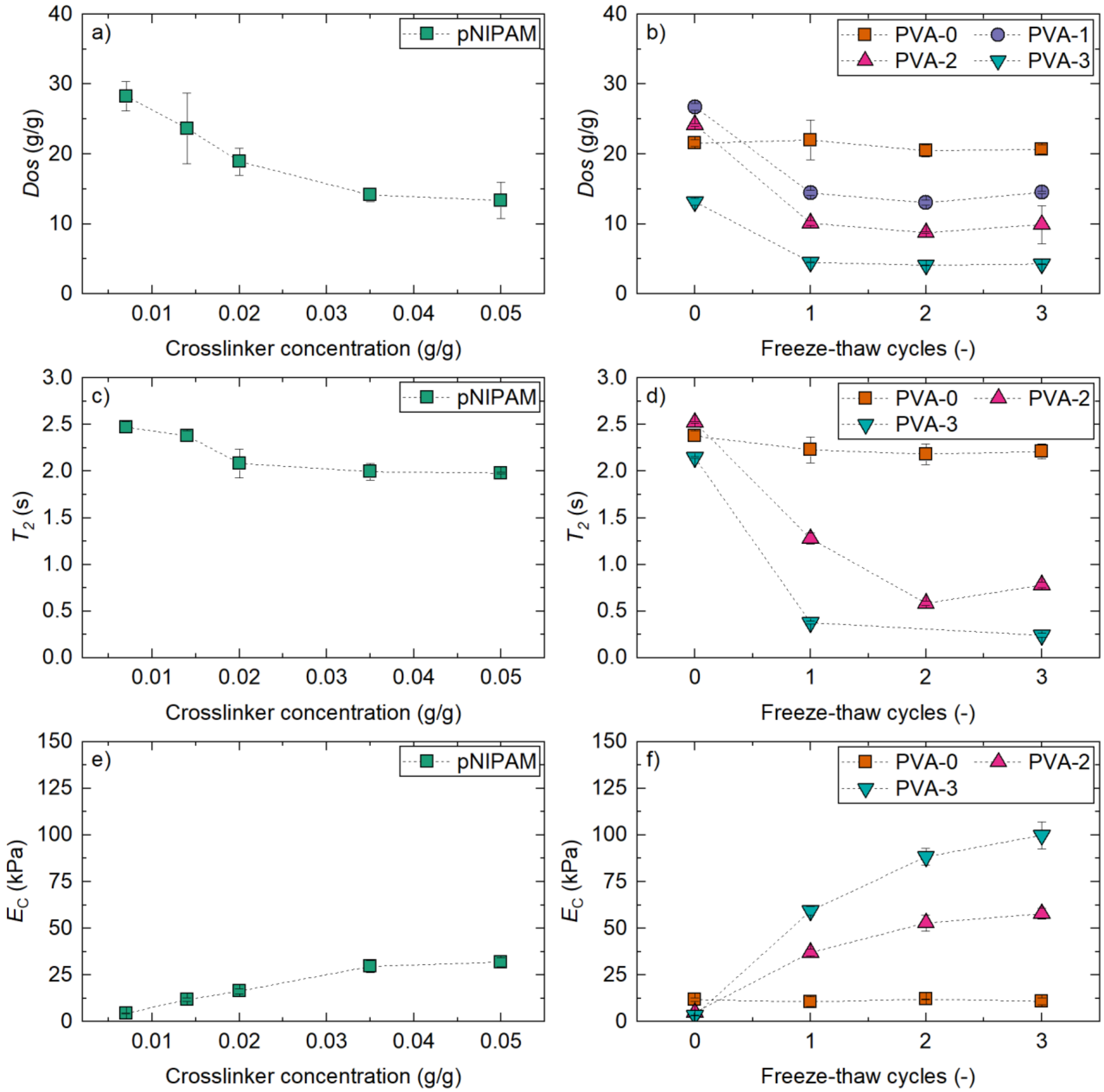

**Figure 1. Swelling degrees (*Dos*) of homopolymers in ethanol (a) and semi-IPNs in 1-butanol (b); $T_2$ relaxation times for homopolymers (c) and semi-IPNs (d) in water; Young's moduli $E_C$ for homopolymers (e) and semi-IPNs (f) in water. The semi-IPN formulations are compared to the PVA-free homopolymer.**

The introduction of PVA into the pNIPAM network to form semi-IPNs increases the swelling behavior of the lyogels (Figure 1b). This trend is also observed in the NMR analysis, where the semi-IPN formulations (PVA-1, PVA-2) exhibit higher relaxation times compared to the homopolymeric formulation with 0% PVA (pNIPAM-2/PVA-0). This effect can be attributed to the hydrophilic nature of PVA, which promotes hydrogen bonding with the solvents used, or increasing pore sizes in the semi-IPNs[19]. These findings align with literature findings on hydrogel swelling, which indicates that increasing PVA content — and consequently the number of hydroxyl groups — enhances the swelling capacity of pNIPAM hydrogels in water[19], [30], [60]. However, at higher PVA concentrations, this effect reverses (PVA-3). Nevertheless, it is important to emphasize that the transferability of these explanations to organic solvents is not straightforward, as swelling behavior depends on the specific solvent and lyogel characteristics. For instance, when comparing butanol-containing lyogels to those in non-polar solvents like butyl acetate, contradictory effects may arise.

The previous results already highlight that the porous structure of the gel network is strongly influenced by the polymers used. In most applications, faster response of the materials are essential, making open-porous networks with large pores a key requirement. Comparing the network structures of homopolymers and semi-IPNs in Figure 2, reveals that the semi-IPN formulations form a broader network, which is expected to facilitate faster mass transport. While homopolymeric pNIPAM gels form small, and spherical pores, semi-IPN form rather irregular structures with larger pores. This effect, known as channel formation in semi-IPN hydrogels, enables faster mass transport into or out of the gel matrix, thereby increasing swelling kinetics[19]. However, increasing PVA concentrations generally reduces pore sizes by forming a denser polymer network, as observed in Figure 2. This suggests that lower PVA concentrations can enhance mass transport kinetics by disrupting the homopolymeric structure of pNIPAM, whereas higher PVA concentrations may inhibit transport processes. These findings align with literature findings that adding PVA to the hydrogel network has the contradictory effects of improving response kinetics and stiffness of the hydrogel matrix but also reducing swelling capacities due to increased stiffness[19], [30], [60], [61].

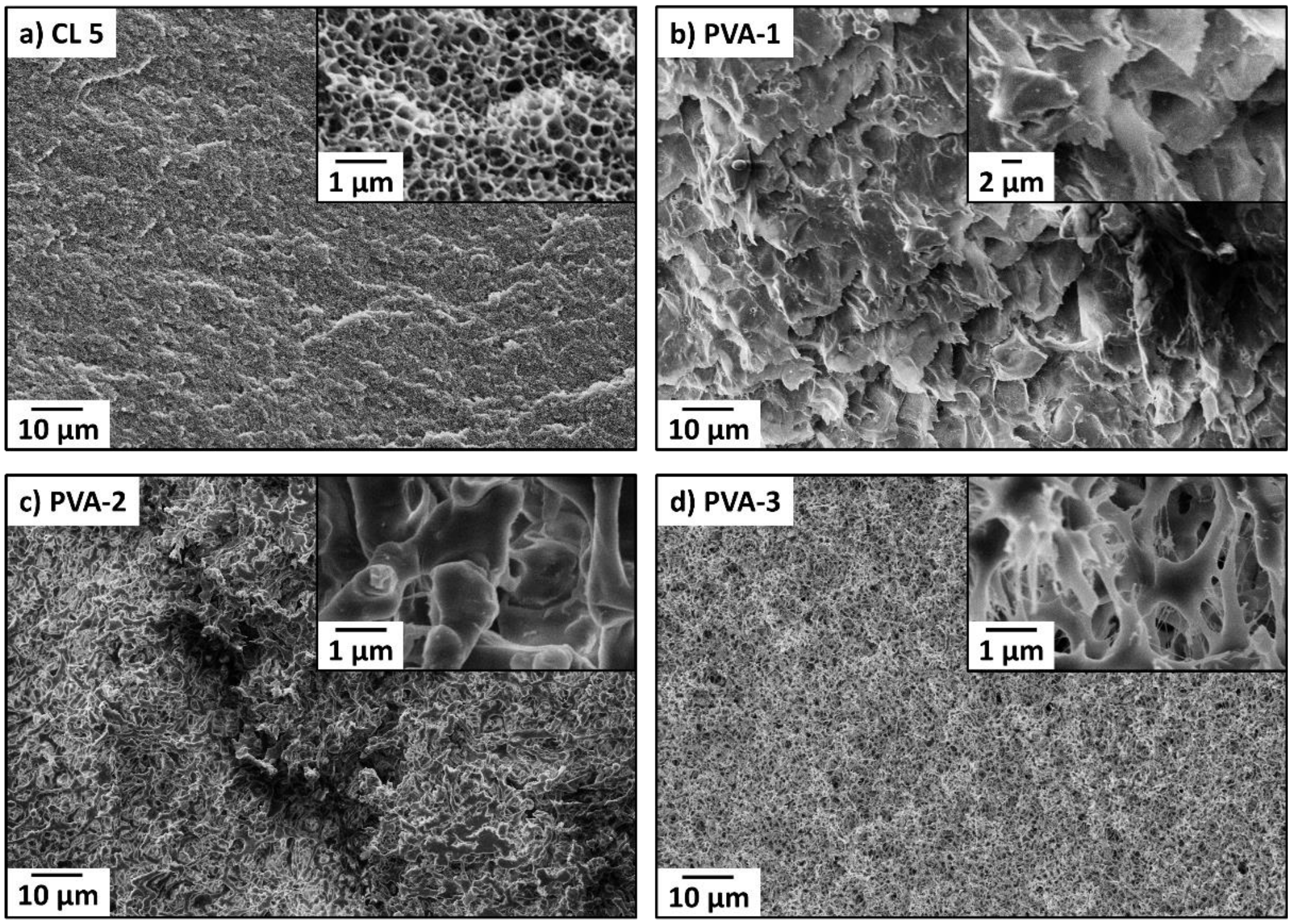

**Figure 2. SEM micrographs of pNIPAM-5 (a) and pNIPAM-PVA semi-IPNs with different PVA concentrations (b-d).**

In conclusion, the swelling capacity of lyogels is strongly influenced by the mechanical and structural properties of the gel matrix, as well as the underlying polymer-solvent interactions. The general finding available in the literature for hydrogels seem to be transferable to organic solvents. SEM analysis suggests that semi-IPNs may enable faster mass transport due to larger pore sizes, potentially leading to shorter response times.

**Influence of Polymer Formulation on Swelling Kinetics**

First, we examine the influence of gel size and crosslinker concentration within the gel matrix of homopolymers on the swelling kinetics. Since the diffusion length of solvent molecules increases with gel size, the diffusion time is longer for larger monoliths, resulting in longer response times for the largest gels (Figure 3 a). As monolith size is a dominant factor, directly comparing gels with varying sizes — resulting from different polymer formulations or solvents — is not feasible. To isolate the influence of polymer formulations on the response kinetics, dimensionless times are used. In this representation, the same formulation exhibits consistent behavior regardless of gel diameter (Figure 3 b), enabling direct comparisons of different polymer formulations independent of gel size and the differing swelling behaviors of the formulations. This difference is evident when comparing Figure 3 b and Figure 3 d, as changes in size have a negligible impact on response behavior, while variations in formulation lead to significant changes.

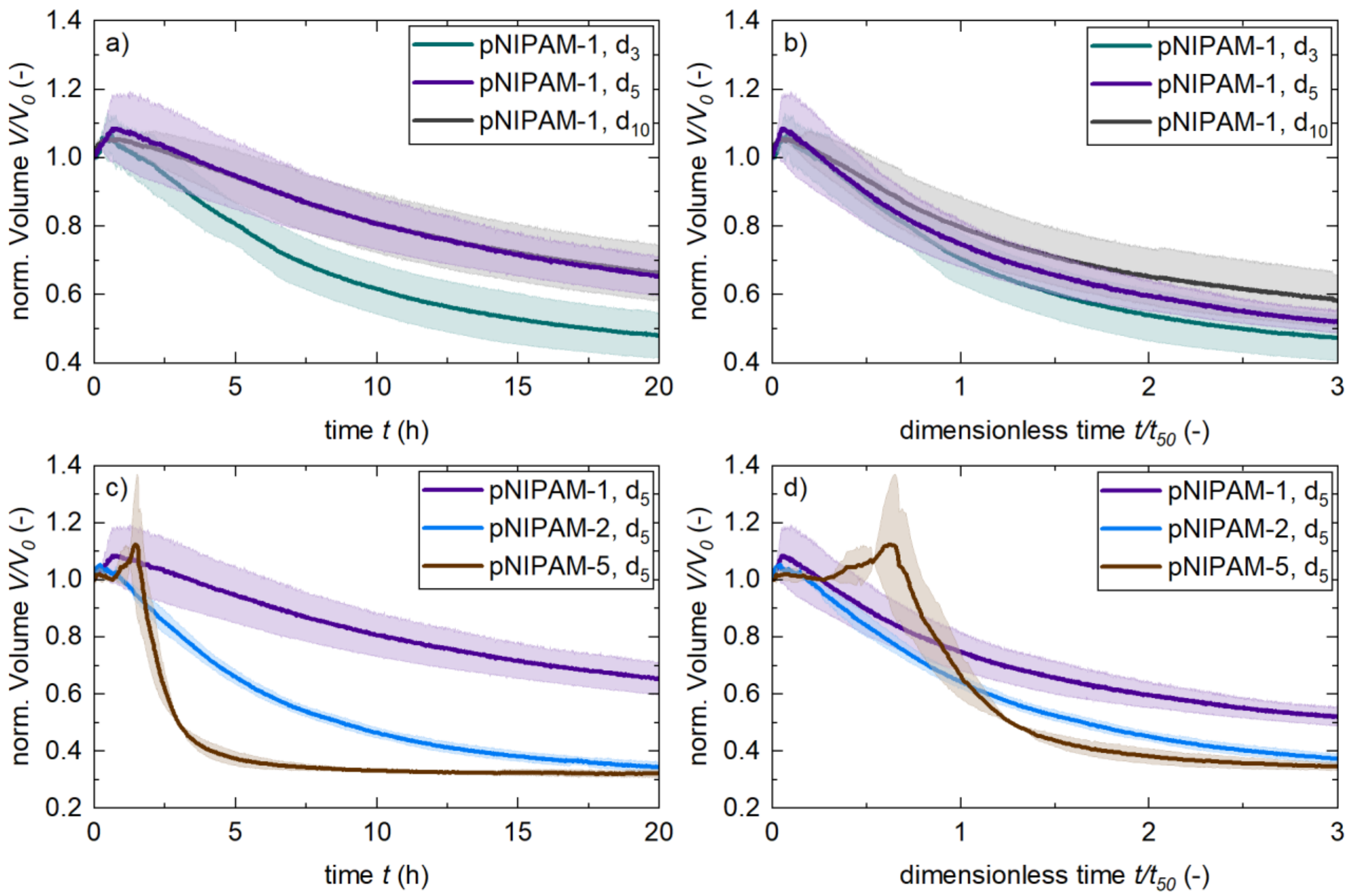

**Figure 3. Solvent exchange kinetics from ethanol to ethyl acetate for pNIPAM lyogels for varying gel diameter (a, b) and crosslinker concentrations (c, d), in absolute (a, c) and dimensionless time representations (b, d).**

Increasing crosslinker concentration facilitates faster mass transport into or out of the lyogel matrix (Figure 3c, d). As noted by Li *et al.*[62] and Zhang *et al.*[56] for hydrogel structures, increased crosslinker concentrations enhance the mechanical stability of the network and leads to smaller, more uniform pore sizes resulting in improved mass transport. However, as noted by Balaguer *et al*.[63], or Diaz-Gonzalez *et al*.[64], excessively high crosslinker concentrations may have a contradictory effect; a significant reduction in pore size can hinder diffusion[64]. In this work, the crosslinker concentrations analyzed fall within the range studied by Zhang *et al*. [56] (0.01-0.12 $mg_{MBA}/mg_{NIPAM}$) for the hydrogel responses. According to our observations, analogous behavior is true for lyogels: increasing crosslinker concentration results in a stiffer matrix without compromising the network's elasticity, thereby preserving its swelling capacity and even improving solvent exchange kinetics (Figure 1).

In the next step, we compare the swelling kinetics of the homopolymers with that of the semi-IPNs — specifically pNIPAM-PVA — and analyze the impact of physical crosslinking due to freeze-thaw methods (Figure 4). When examining the effect of PVA content on swelling kinetics, two distinct effects emerge.

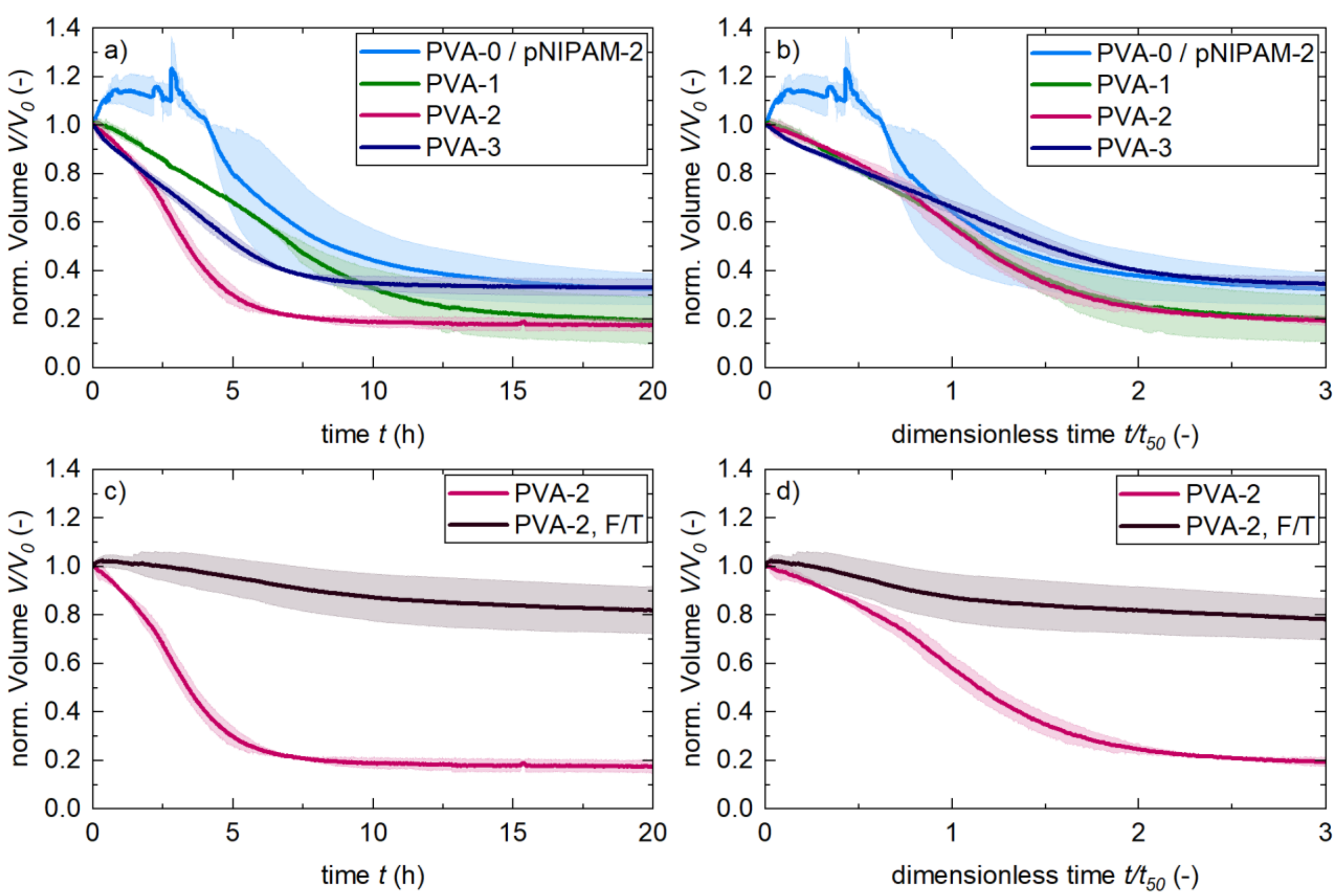

**Figure 4. Swelling kinetics of pNIPAM-PVA (semi-IPN, $d_5$) in a solvent exchange step from 1-butanol to butyl acetate (a, b) and the comparison of PVA-2 without and with three applied freeze-thaw sequences (c, d) in absolute (a, c) and dimensionless time representations (b, d).**

First, the addition of small amounts of PVA to the lyogel structure significantly enhances the shrinkage kinetics compared to the homopolymers. The incorporation of a copolymer to create a (semi-)IPN of is known to form channels within the gel structure, which in case of hydrogels improve the mass transport properties in temperature-induced swelling , as previously described by Haq *et al*.[19]. From Figure 4, it can be derived that this knowledge can be transferred to lyogel behavior in solvent-induced swelling. Further, these findings align with the observed effect of increased crosslinker content in the homopolymers enabling faster mass transport. PVA acts as a structural reinforcement in these formulations, making the gel more porous and less compact, similar to the effect of increased crosslinker concentrations observed in the homopolymers. In both systems, the increasing porosity facilitates faster mass transport, thereby improving swelling kinetics.

However, further increasing the PVA concentration leads to reduced swelling kinetics. This may be due to a reduction in the swelling capacity and elasticity of the gels. These effects are likely a result of stronger interactions between PVA chains, which form crystalline regions within the network (Figure 2). This leads to denser network structures, reducing the free volume available for solvent diffusion. In comparison to the homopolymers, increasing the PVA concentration to 1% and 2% (PVA-1, PVA-2) enhances mass transport, whereas at 4% (PVA-3), it already surpasses the optimal range, leading to reduced response kinetics of the lyogels.

While the PVA integration introduces additional physical crosslinking points to the network, freeze-thaw cycles further increase the crystalline regions in the lyogels, thereby amplifying the physical crosslinking[65], [66]. The formation of crystalline regions decreases the hydrophilicity of the gel, as observed at 4% PVA (Figure 1c, d), resulting in lower swelling capacity and smaller pores in the gel network. Consequently, the network density increases, further restricting solvent diffusion and reducing swelling kinetics. This assumption is supported by Figure 2, and Figure 4 c, which compares the impact of freeze-thaw cycles on response kinetics. In this analysis, a significant reduction in shrinkage and extended response times was observed in these gels. In contrast to the moderate increase in crosslinker concentration, analyzed in the pNIPAM homopolymers, or the low PVA concentrations in the semi-IPNs, the freeze-thaw cycles result in more extensive crosslinking. As a result, these cycles do not improve transport processes; instead, they lead to significantly stiffer networks with reduced swelling capacity and decreased swelling kinetics. Overall, although the freeze-thaw method creates a stiffer network and improves the handling of the gels, it significantly reduces the gels' ability to exhibit strong swelling effects.

These results emphasize that the network structure of the gels plays a crucial role in their swelling behavior and, consequently, requires detailed adaptation to potential applications. The network structure is influenced by factors such as variations in crosslinker concentration, the creation of a (semi-)IPN, and post-processing techniques like freeze-thaw cycles, all of which significantly affect swelling kinetics and capacities in organic solvents.

**Influence of Solvent Nature on Swelling Kinetics**

In the previous section, it was shown that the response kinetics of the lyogels are strongly influenced by their porous structure. However, all formulations exhibit similar volumetric changes as they are analyzed in the same solvents. Thus, this section focuses on the influence of solvents on the response kinetics of pNIPAM, examining the effect of different solvent exchange systems containing of carboxylic acids, alcohols and carboxylic esters.

Generally, hydrogel shrinkage follows exponential or power-law kinetics[25], [29], [44]. However, at the onset of the shrinkage process — before general contraction begins — an initial increase in gel volume is occasionally observable.

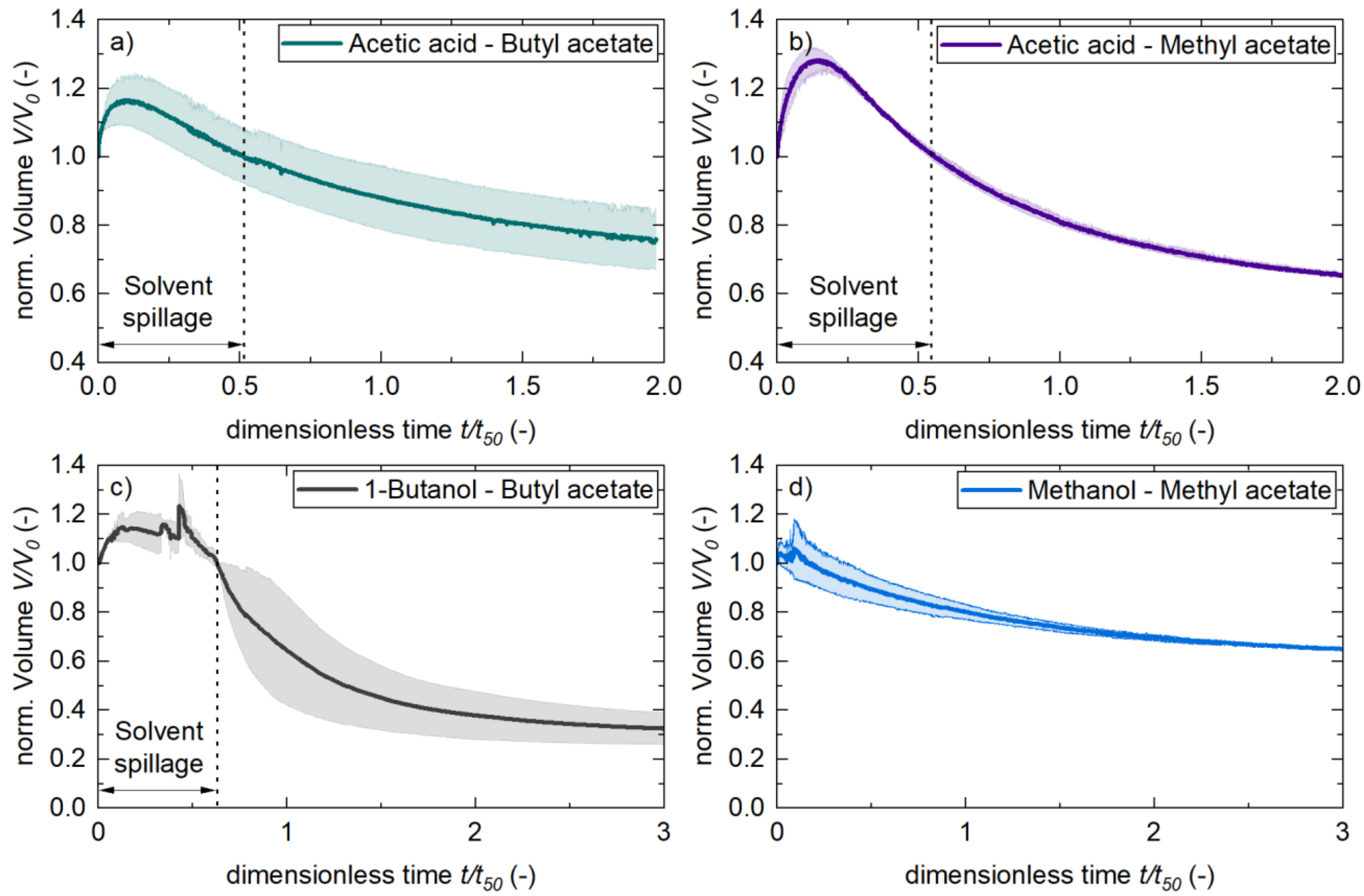

**Figure 5. Swelling kinetics during the solvent exchange of homopolymeric pNIPAM gels ($d_{10}$) in different organic solvents: solvent exchange from a) acetic acid to butyl acetate, b) acetic acid to methyl acetate, c) 1-butanol to butyl acetate, and d) methanol to methyl acetate.**

The initial swelling of the lyogels at the start of the shrinkage process may be attributed to the solvent spillage effect as described by Bueno *et al.*[50]. In the first step, the surrounding liquid diffuses into the gel matrix, while the other solvent begins to diffuse out[50]. This solvent uptake leads to pressure accumulation within the gel matrix until a certain threshold is reached, at which point the gel releases the accumulated solvents[50]. This behavior is especially pronounced during the solvent exchange from acetic acid to carboxylic esters, particularly with methyl acetate. In these systems, it appears that the lyogel matrix can retain a larger amount of liquid before spillage occurs. Since solvent spillage is more pronounced in the exchange from acetic acid to methyl acetate, compared to butyl acetate, the molecular size of the solvents may be an influencing factor. Methyl acetate and methanol are smaller and exhibits faster diffusion than 1-butanol and butyl acetate[67], [68]. These findings align with recent work by Takeshita *et al.* [51], who compared the solvent exchange process during aerogel production and noted that the initial swelling before shrinkage was more pronounced in methanol than in ethanol. However, solvent spillage was observed during the solvent exchange from 1-butanol to butyl acetate but not from methanol to methyl acetate, highlighting that differences in swelling behavior of the gel among the solvents also contribute to this effect. Additionally, the distinct volumetric changes measured during the initial swelling phase in different processes suggest that factors beyond diffusion properties, such as varying intermolecular interactions in different systems, further influence the swelling and shrinkage mechanisms.

Although no clear increase in volume can be measured during the solvent exchange from alcohols to carboxylic esters (see Figure 5 and Figure 4), it is evident that the volume change does not follow typical power-law kinetics. In all systems, the curves begin with either a constant or increasing volume, rather than the expected strong volume decrease commonly observed in hydrogels[25], [29], [39], [44]. Possible reasons include the relatively slower diffusion of organic solvents compared to water[69], attributable to differences in molecular size[67], [68]. Another contributing factor is the bidirectional flow of solvents: one solvent diffusing into the matrix while another is expelled, which is most prominent in the solvent spillage effect.

In this study, the longer response times of several days allowed for a more straightforward comparison of the swelling and shrinkage mechanisms occurring in the different solvents. However, for potential applications, these findings need to be compared and optimized in conjunction with the results from the previous chapter.

## CONCLUSIONS

This study demonstrates promising methods for tailoring lyogel formulations by optimizing their solvent-induced swelling behavior, advancing beyond the previous focus on hydrogel systems. The first part focused on characterizing the lyogels to assess the impact of crosslinking and copolymer inclusion on gel structure and properties. Solvent affinity (via NMR relaxometry), swelling capacity, mechanical behavior, and microstructure were evaluated across the different formulations. The second part examined the swelling kinetics of pNIPAM homopolymers and semi-IPNs, alongside key lyogel characteristics, such as swelling degree and mechanical properties. In this study, it was demonstrated that the degree of crosslinking and the presence of copolymers significantly influence not only the swelling degree but also the overall polymer network, thereby affecting the swelling kinetics of pNIPAM lyogels. In contrast to prior studies addressing aqueous solutions, our analysis was conducted in organic solvents and their mixtures, and we investigated the impact of solvent nature on solvent-induced swelling kinetics. Solvent-specific effects highlight the importance of molecular size and diffusion properties in solvent exchange processes. Notably, the solvent spillage effect was most pronounced in systems with smaller, faster-diffusing molecules, underscoring the role of solvent properties on mass transport effects in solvent-induced swelling.

Overall, our findings pave the way for tailoring the network structure of lyogels for potential applications, optimizing mechanical stability while preserving strong swelling behavior and rapid kinetics to enhance practical performance.

## Acknowledgment

This project is funded by the Deutsche Forschungsgemeinschaft (DFG, German Research Foundation) – SFB 1615 – 503850735. The authors thank the Electron Microscopy Unit (BEEM) for access to SEM.